\documentclass[aps,pre,preprint,groupedaddress]{revtex4-1}

\usepackage{color}
\usepackage{amsmath}
\usepackage{graphicx}
\usepackage{amssymb}

\begin{document}


\title{Blowup as a driving mechanism of turbulence in shell models}
\thanks{This work was supported by CNPq under grants 477907/2011-3 and 305519/2012-3.}


\author{Alexei A. Mailybaev}
\affiliation{%
Instituto Nacional de Matem\'atica Pura e Aplicada -- IMPA, Rio de Janeiro,
Brazil
\footnote{Estrada Dona Castorina 110, 22460-320 Rio de Janeiro, RJ, Brazil. 
E-mail: alexei@impa.br}
}%


\date{\today}

\begin{abstract}
Since Kolmogorov proposed his phenomenological theory of hydrodynamic turbulence in 1941, 
the description of mechanism leading to the energy cascade 
and anomalous scaling remains an open problem in fluid mechanics. Soon after, in 1949 Onsager noticed that the scaling properties in inertial range imply non-differentiability of the velocity field in the limit of vanishing viscosity. This observation suggests that the turbulence mechanism may be related to a finite-time singularity (blowup) of incompressible Euler equations. However, the existence of such blowup is still an open problem too. In this paper, we show that the blowup indeed represents the driving mechanism of inertial range for a simplified (shell) model of turbulence. Here, blowups generate coherent structures (instantons), which travel through the inertial range in finite time and are described by universal self-similar statistics. The anomaly (deviation of scaling exponents of velocity moments from the Kolmogorov theory) is related analytically to the process of instanton creation using the large deviation principle. The results are confirmed by numerical simulations. 
\end{abstract}


\maketitle

\section{Introduction}

Describing the mechanism of developed turbulence for the 3D Navier-Stokes equations 
remains an important open problem in fluid mechanics. 
It encompasses various questions, and in this work 
we address the anomalous statistics of velocity moments in inertial range 
and the dissipation 
anomaly (existence of finite dissipation in the inviscid limit).
These questions remain the hot research topic since Kolmogorov presented the 
phenomenological theory of inertial range in 1941~\cite{kolmogorov1941local}. 
This theory of isotropic homogeneous turbulence leads to the power-law 
dependence of velocity moments on spatial scales, providing  
the scaling exponents $\zeta_p = p/3$ obtained on dimensional grounds.  
The exact scaling exponents deviate from the Kolmogorov theory.  
These deviations, called the anomalous corrections, are universal and become large with increasing $p$. 
Though a lot of knowledge 
is available now on the described anomalous phenomena, their mechanism is still not well understood~\cite{frisch1995turbulence,cardy2008non}. 

In 1949, Onsager~\cite{eyink2006onsager} related scaling properties of turbulent flow 
in inertial range with the regularity of solutions obtained in the limit of vanishing viscosity. He conjectured that the anomalous turbulent dissipation requires the limiting velocity field to be non-differentiable with the H\"older continuity exponent $h \le 1/3$. This conjecture was proved later~\cite{eyink1994energy,constantin1994onsager}. Irregularity of inviscid solutions allows considering the flow as a multifractal set with a continuous infinity of dimensions~\cite{aurell1992multifractal,frisch1995turbulence}, which explains the nonlinear shape of scaling exponents $\zeta_p$. Development of the theory of turbulence in this way has the fundamental obstacle. It is the problem of blowup, i.e., the formation of a finite-time singularity in the incompressible 3D Euler equations from smooth initial data of finite energy. 
So far, the existence of blowup remains an open problem~\cite{gibbon2008three}.

Simplified models help in understanding the turbulence phenomena. 
In this respect, the  Gledzer--Ohkitani--Yamada (GOY) shell model of 
turbulence~\cite{gledzer1973system,ohkitani1989temporal} was successful in describing 
several nontrivial properties including the inertial range with anomalous dissipation and scaling. 
Shell models represent the dynamics in terms of characteristic (shell) velocities corresponding 
to a discrete set of wavenumbers increasing in geometric progression, and allow 
reliable numerical simulation at very high Reynolds numbers. 
The Sabra shell model proposed in~\cite{l1998improved} is characterized by improved 
regularity in the inertial range. Despite of large effort~\cite{biferale2003shell}, 
the theory of turbulence for shell models, which would follow directly from the model 
equations and describe the observed statistics, is not yet accessible. 
On the other hand, the problem of blowup was recently  formalized~\cite{constantin2007regularity} 
and understood~\cite{dombre1998intermittency,mailybaev2012c}. 
The blowup in the Sabra shell model has self-similar universal structure~\cite{mailybaev2012}. 
Possible relation of such structure to the statistics of turbulence was 
discussed in~\cite{gilson1997towards,l2001outliers}. Cascade models of turbulence~\cite{eggers1991does} represent 
the extended version of shell models, where each shell is described by a large 
(though fixed) number of variables. These models lead to anomalous intermittent 
dynamics~\cite{uhlig1997local}, and the universal self-similar blowup 
was observed numerically in 
the inviscid cascade model~\cite{uhlig1997singularities}. See also~\cite{siggia1978model,
nakano1988,l2001outliers,l2002quasisolitons} for other numerical observations of 
self-similar blowup 
in shell models.    

In this paper, we establish a direct link between the blowup 
and the turbulent dynamics in inertial range for the Sabra shell model. 
We show that the blowup-like structures 
dominate the turbulent fluctuations and can be described as a ``gas'' of 
instantons. 
The instantons (coherent structures of shell velocities, which traverse the inertial range in direction of large wavenumbers) 
are represented and analyzed in terms of velocity local maxima. 
The striking property of instantons is that they maintain the universality and self-similarity 
of blowup, though with slightly different scaling exponents and in statistical sense. 
This statistical universality  of instantons was observed earlier in~\cite{mailybaev2012computation}.
Then we show that instanton creation is the main process 
driving the inertial range dynamics. 
This allows deriving the probability density function (PDF) for 
instanton amplitudes explicitly in terms of the anomalous scaling exponents $\zeta_p$ 
by using the large deviation principle. The obtained results fully agree with numerical simulations 
and are also confirmed analytically for a class of instanton creation models. 
Finally, we discuss some qualitative changes in the turbulent regime,
which occur with a change of model parameter.  

The paper is organized as follows. Section~\ref{sec2} introduces the Sabra shell model. 
The blowup universal properties in the inviscid model are described in Section~\ref{sec3}. 
In Section~\ref{sec4} we consider statistics of maxima of velocity amplitudes 
and introduce a way to identify the instantons. 
Section~\ref{sec5} describes the universal self-similar statistics of instantons. 
In Section~\ref{sec6}, we find universal expressions for PDFs 
of instantons using the large deviation principle. Section~\ref{sec7} presents the analytical theory for a specific class of instanton creation models. Section~\ref{sec8} describes a different turbulent regime, which is dominated by a single blowup. The results are summarized in Section~\ref{sec9}.

\section{Model}
\label{sec2}

In shell models of turbulence, the Fourier space is represented by a series of shells $n = 0,1,2,\ldots$ corresponding to wavenumbers $k_n = \lambda^n$ with $\lambda = 2$.
We consider the Sabra shell model~\cite{l1998improved}
\begin{equation}
\begin{array}{rl}
\displaystyle
\frac{du_n}{dt} 
\displaystyle
= & i[k_{n+1}u_{n+2}u_{n+1}^*-(1+c)k_{n}u_{n+1}u_{n-1}^*
\\[5pt] &\displaystyle
-ck_{n-1}u_{n-1}u_{n-2}]-\nu k_n^2u_n+f_n,
\end{array}
\label{eq1}
\end{equation} 
where $u_n$ is the complex shell velocity, which can be understood as the Fourier component 
of the velocity field at the shell wavenumber $k_n$, 
$\nu \ge 0$ is the viscosity, and $c$ is the parameter controlling nonlinear coupling of the shells. 
The terms $f_n$ model external forces at large scales and, thus, 
they are usually restricted to the first few shells. 
The inviscid system with no forcing ($\nu = f_n = 0$) conserves the energy $E = \frac{1}{2}\sum_n|u_n|^2$. 
The second quadratic invariant $H = \sum_n c^{-n}|u_n|^2$ is associated with 
the helicity for $c = -0.5$ when $c^{-n} = (-1)^nk_n$. 
Additionally, there are four symmetry transformations 
\begin{equation}
t \mapsto t-t_0;
\label{eqSymn}
\end{equation}
\begin{equation}
u_n \mapsto e^{i\theta_n}u_n,\quad
\theta_{n} = \theta_{n-1}+\theta_{n-2};
\label{eq2}
\end{equation}
\begin{equation}
t \mapsto t/a,\quad u_n \mapsto au_n; 
\label{eq15n}
\end{equation}
\begin{equation}
u_n \mapsto \lambda u_{n+1}.
\label{eqSym2n}
\end{equation}
Here Eqs.~(\ref{eqSymn}) and (\ref{eq2}) can be associated with the time and physical space translations, 
while Eqs.~(\ref{eq15n}) and (\ref{eqSym2n}) correspond to the time and space scaling, see~\cite{biferale2003shell}. 

\section{Blowup in inviscid model}
\label{sec3}

Let us consider solutions $u_n(t)$ with the finite norm $|u|_1 < \infty$ defined as
\begin{equation}
|u|_1 = \left(\sum_nk_n^2|u_n|^2\right)^{1/2}.
\label{eq_norm}
\end{equation}
The blowup represents a singularity given by
\begin{equation}
|u|_1 \to \infty \quad \textrm{as}\quad t \to t_c^-,
\label{eq_sin}
\end{equation}
which develops in finite time $t_c < \infty$ from initial condition of 
finite norm~\cite{constantin2007regularity}. Note that the singularity is described by the norm, 
while each particular shell speed $u_n(t)$ remains finite and smooth. 
This reflects the fact that the shell model corresponds to dynamics in the Fourier space, where the condition like (\ref{eq_sin}) implies the divergence of velocity derivatives in physical space, i.e., infinite vorticity. The blowup is only possible in the inviscid shell model, and the uniqueness of solution is insured only for $t < t_c$~\cite{constantin2007regularity}. 

Let us consider the inviscid model with vanishing forcing terms, $\nu = f_n = 0$. Then we write Eq.~(\ref{eq1}) as
\begin{equation}
\frac{du'_n}{dt} = N_n[u'],\quad
u'_n = ik_nu_n,
\label{eq1inv}
\end{equation} 
with the quadratic nonlinearity
\begin{equation}
N_n[u'] = -\lambda^{-2}u'_{n+2}u_{n+1}^{\prime *}+(1+c)u'_{n+1}u_{n-1}^{\prime *}
-c\lambda^2u'_{n-1}u'_{n-2}.
\label{eqN}
\end{equation} 
Following the approach suggested by Dombre and Gilson~\cite{dombre1998intermittency} (see also \cite{mailybaev2012}), we consider the renormalized time $\tau$ and shell speeds $w_m$ introduced as
\begin{equation}
\begin{array}{rcl}
t & = & \displaystyle 
t_0+\int_0^\tau \exp\left[-\int_0^{\tau'} A(\tau'') d\tau''
\right]d\tau',
\\[17pt]
u'_n & = & \displaystyle 
\exp\left[\int_0^\tau A(\tau') d\tau'\right]w_n,
\end{array}
\label{eqS.6}
\end{equation}
where $\tau = 0$ corresponds to the initial time $t_0$, and $A(\tau)$ is specified below. It is
straightforward to check that
\begin{equation}
\frac{dw_n}{d\tau} = N_n[w]-Aw_n,
\label{eqS.5}
\end{equation}
where $N_n[w]$ has the form (\ref{eqN}) written in terms of $w_n$ instead of
$u'_n$. One can also check that Eq.~(\ref{eqS.5}) conserves the sum $\sum |w_n|^2$ if we choose
\begin{equation}
A(\tau) = \mathrm{Re}\sum_n w_n^*N_n[w]\Big/\sum_n |w_n|^2.
\label{eqS.5b}
\end{equation}

The idea of the above transformation is that Eq.~(\ref{eqS.5}) admits an  asymptotic traveling wave solution 
of the form~\cite{dombre1998intermittency}
\begin{equation}
w_n(\tau) = W(n-s\tau),
\label{eqTW}
\end{equation}
where $s$ is the wave speed and $W(\xi)$ is a function vanishing as $\xi \to \pm\infty$. 
This traveling wave exists for a large range of shell model parameter $c$, and 
it is determined up to symmetries induced by Eqs.~(\ref{eqSymn})--(\ref{eqSym2n}). 
For the original shell speeds $u_n(t)$ related to $w_n(\tau)$ by Eqs.~(\ref{eq1inv}) 
and (\ref{eqS.6}), solution (\ref{eqTW}) 
yields~\cite{dombre1998intermittency,mailybaev2012} 
\begin{equation}
u_n(t) = -ik_n^{-y_0}U(k_n^{z_0}(t-t_c)),
\label{eq21}
\end{equation}
where 
\begin{equation}
U(t-t_c) = \exp\left[\int_0^\tau A(\tau')d\tau'\right]W(-s\tau),
\label{eq22}
\end{equation}
\begin{equation}
z_0 = \frac{1}{\log \lambda}\int_0^{1/s}A(\tau) d\tau, \quad y_0 = 1-z_0,
\label{eq17}
\end{equation}
\begin{equation}
t_c = t_0+\int_0^\infty \exp\left[-\int_0^{\tau'} A(\tau'') d\tau''
\right]d\tau'.
\label{eq19}
\end{equation}
If $y_0 < 1$, then Eq.~(\ref{eq21}) describes the asymptotic form of 
blowup at finite time $t_c < \infty$. 
In this asymptotic form, $y_0$ is the universal scaling exponent independent of initial conditions, 
and the function $U(t)$ describes the universal self-similar shape of the blowup 
given up to the scaling symmetry of the Sabra model. The equality $y_0+z_0 = 1$ reflects the dimensional relation $t_n-t_c \propto (v_nk_n)^{-1}$, where $v_n = \max_t|u_n(t)|$ and $t_n$ is the corresponding time.     
For details of the derivations and the rigorous theory, which associates the  
traveling wave (\ref{eqTW}) with a fixed-point attractor of the Poincar\'e map, see \cite{mailybaev2012c}. 

\begin{figure}
\centering \includegraphics[width = 0.7\textwidth]{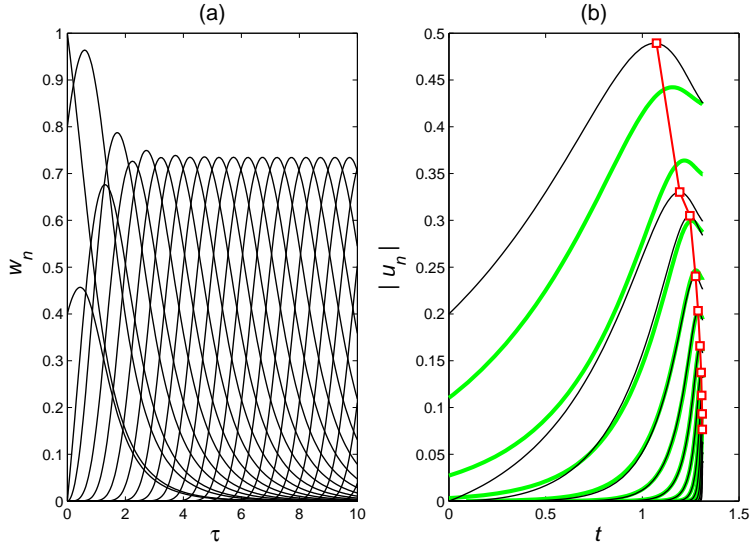}
\caption{(Color online) (a) Traveling wave formation in the dynamics of renormalized inviscid Sabra model. 
Shown are the curves $w_n(\tau)$ with $n = 0,1,\ldots$ increasing from the left to the right. 
(b) The corresponding dynamics of shell speeds $u_n(t)$ for $n = 2,3,\ldots$. 
Bold green (light gray) curves show the universal self-similar asymptotic form of blowup. 
Red squares indicate the correlated sequence of maxima $v_n = \max_t|u_n(t)|$.}
\label{fig1}
\end{figure}

As an example, let us consider the case $c = -0.5$. Solution $w_n(\tau)$ of the renormalized 
system (\ref{eqS.5}) for real initial conditions is shown in Fig.~\ref{fig1}a. 
One can clearly see the formation of traveling wave solution (\ref{eqTW}). 
Solution for the original shell speeds $u_n(t)$ is presented in Fig.~\ref{fig1}b, 
which blows up at finite time $t \to t_c$ given by Eq.~(\ref{eq19}). 
Using Eqs.~(\ref{eq22}) and (\ref{eq17}), 
we compute the scaling exponent $y_0 = 0.281$ and 
the function $U(t)$. 
The bold green curves in Fig.~\ref{fig1}b show the asymptotic self-similar solution (\ref{eq21}) 
for the blowup, and one can readily see the convergence.
Numerical analysis confirms asymptotic stability of the traveling wave solution in 
Fig.~\ref{fig1}a due to both real and complex perturbations. 
As we already mentioned, this implies that Eq.~(\ref{eq21}) provides the universal asymptotic 
form of blowup. 

\begin{figure}
\centering \includegraphics[width = 0.7\textwidth]{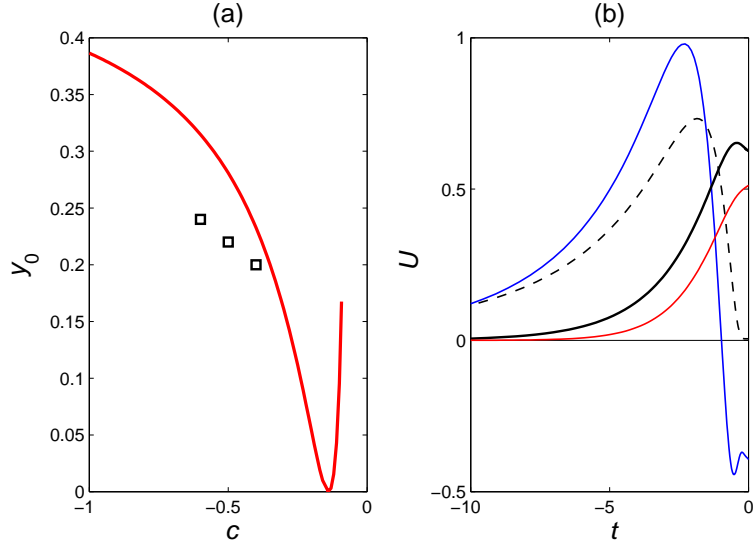}
\caption{(Color online) (a) Dependence of the blowup scaling exponent $y_0$ on the Sabra model parameter $c$. 
Squares show the scaling exponents $y$ of instantons.
(b) The universal function $U(t)$ in the asymptotic expression (\ref{eq21}) for $c = -0.1$ 
(upper curve, blue), 
$c = -0.139$ (dotted curve), $c = -0.5$ (solid black curve) and $c = -1$ (lower curve, red).}
\label{fig2}
\end{figure}

Similar traveling wave solutions exist for $c < -0.092$. The corresponding scaling 
exponent $y_0$ and function $U(t)$ are shown in Fig.~\ref{fig2}. 
The function $U(t)$ is monotonous for $c < -0.671$, possesses a single extremum (maximum) 
for $-0.671 < c < -0.139$, and has several extrema for $-0.139 < c < -0.092$. 
At $c = -0.139$, we have $U(0) = 0$ and the scaling exponent attains the minimum $y_0 = 0$. 
This fact can be understood using the energy conservation argument. 
Indeed, $y_0$ in Eq.~(\ref{eq21}) cannot be negative, otherwise the shell speeds and the energy would grow infinitely. 
In the case $y_0 = 0$ all the energy is transported to large shells as $t \to t_c^-$, 
so that no energy remains in each shell at the time of blowup, i.e., $U(0) = 0$.

The real traveling wave solution (\ref{eqTW}) becomes unstable with respect to complex perturbations 
at the critical value $c = -0.092$. For $c > -0.092$ analysis of the blowup requires more 
sophisticated techniques, see~\cite{mailybaev2012c}, which is beyond the scope of this paper. 
  
\section{Instantons in inertial range of turbulent regime}
\label{sec4}

It is known that, for the parameter $c = -0.5$, the Sabra model with small viscosity 
(large Reynolds number) demonstrates chaotic intermittent behavior. 
Statistical properties of this system have much in common with 
the developed turbulence of the 3D Navier-Stokes equations~\cite{l1998improved}. 
In particular, it possesses a wide (increasing with the Reynolds number) inertial range of wavenumbers $k_n$ 
separating the scales influenced by forcing (small $k_n$) 
and the scales dominated by viscosity (large $k_n$). 
This inertial range is responsible for the energy cascade, i.e., 
to the flux of energy produced in the forcing range by external forces 
to the viscous range, where it is dissipated due to viscosity. 
Existence of a positive limit of mean dissipation rate for infinite Reynolds numbers 
constitutes the famous dissipation anomaly of turbulent hydrodynamic flows. 

\begin{figure}
\centering \includegraphics[width = 0.5\textwidth]{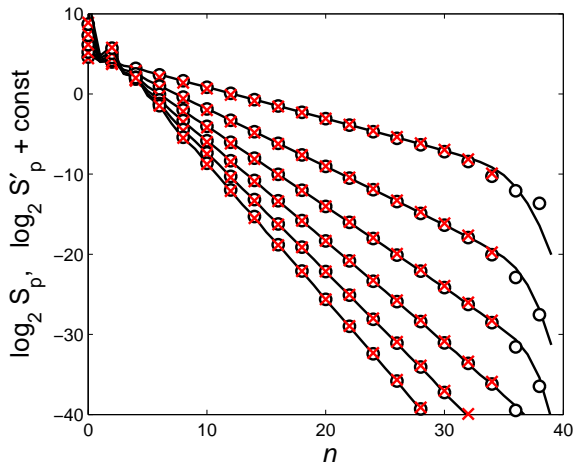}
\caption{(Color online) Solid black lines present the velocity moments $S_p(k_n) = \langle |u_n|^p\rangle$ 
for $p = 1,\ldots,6$. Black circles determine the functions $S'_p(k_n)$ 
from Eq.~(\ref{eqM2}); for better comparison of slopes, the graphs are shifted in vertical direction and only even $n$ are shown. 
Red crosses show similar functions $S'_p(k_n)$ computed for the local maxima $v_n$ 
corresponding to stable instantons only. Three types of structure functions determine 
equal slopes in the inertial range given by the scaling exponents $-\zeta_p$.}
\label{fig3}
\end{figure}

The important quantitative characteristic of inertial range is given by the structure functions (velocity moments). In the inertial range, these functions depend on $k_n$ as power laws
\begin{equation}
S_p(k_n) = \langle |u_n|^p \rangle \propto k_n^{-\zeta_p}.
\label{eqM}
\end{equation}
In this expression $p$ is an arbitrary real number; traditionally, the computations are carried out for positive integer values of $p$. 
The scaling exponents $\zeta_p$ are universal, i.e., they are independent both of the forcing and viscosity. 
Figure~\ref{fig3} presents the functions $S_p(k_n)$ in logarithmic coordinates for the Sabra model with $c = -0.5$. 
These results are based on direct numerical simulation of Eq.~(\ref{eq1}) with $40$ shells, 
viscosity $\nu = 10^{-14}$ and the constant forcing at the first two shells, $f_0 = 1+i$ and $f_1 = f_0/2$. 
One can clearly distinguish the forcing range corresponding roughly to the shells $n \le 5$, the viscous range of shells $n \ge 32$, and the linear part in between indicating the inertial range. 

\begin{figure}
\centering \includegraphics[width = 0.5\textwidth]{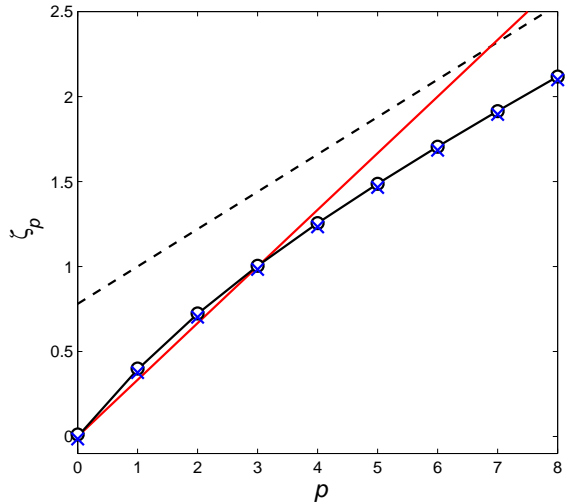}
\caption{(Color online) Anomalous scaling exponents $\zeta_p$ computed for the velocity moments $S_p$ (black line) and for the functions $S'_p$ (circles and crosses correspond to the sums over all maxima and over maxima from stable instantons, respectively). 
The red (gray) line $\zeta_p = p/3$ corresponds to the phenomenological K41 theory. The dotted line shows the upper bound (\ref{eq20}) based on the instanton scaling.}
\label{fig4}
\end{figure}

The phenomenological theory developed by Kolmogorov (K41~\cite{frisch1995turbulence,kolmogorov1941local}) predicts the linear dependence $\zeta_p = p/3$ for the scaling exponents. However, the exact scaling exponents $\zeta_p$ depend nonlinearly on $p$. This deviation from the K41 theory is called the anomaly. The scaling exponents are presented in Fig.~\ref{fig4}. The two exact values of scaling exponents are known. The first one is $\zeta_0 = 0$ since $|u_n|^0 = 1$. The second exact exponent is $\zeta_3 = 1$, which is a necessary condition for the dissipation anomaly, see, e.g.,~\cite{l1998improved}. The scaling exponents of the 3D Navier--Stokes turbulence are close to the ones given by the Sabra model~\cite{frisch1995turbulence,l1998improved}.
  
In this section we establish a link between the anomalous turbulent statistics 
and the blowup phenomenon for the Sabra shell model. 
The blowup analysis of the inviscid model 
is relevant in the inertial range, where viscosity is insignificant. However, there is an essential difference related to initial conditions. For the blowup considered in Section~\ref{sec3}, finiteness of the norm (\ref{eq_norm}) requires decay of initial shell speeds faster than $k_n^{-1}$. This condition is violated in the inertial range of developed turbulence, which is characterized by the power-law decay (\ref{eqM}) with $\zeta_1 \approx 0.39$. We will see that this difference leads to the transformation of the blowup with universal self-similar asymptotic form to coherent structures with universal self-similar  statistics. 

Identification of these coherent structures in turbulent regime is strongly facilitated, if we consider local maxima $v_n = \max_t |u_n(t)|$ of shell speed amplitudes. An extra subscript is necessary to index all the local maxima in shell $n$, but we will drop it for the sake of simplicity of notations. The new ``structure'' functions are defined as
\begin{equation}
S'_p(k_n) = \frac{1}{Tk_n}\sum v_n^{p-1},
\label{eqM2}
\end{equation} 
where the sum is taken over all local maxima $v_n$ observed for the speed amplitude $|u_n(t)|$ during a large time interval $0 \le t \le T$. 
By a simple dimensional consideration, one finds that each local maximum $v_n$ has the characteristic time $\Delta t_n \sim (k_nv_n)^{-1}$ determining the time interval, where $|u_n(t)| \sim v_n$. For the velocity moment $\langle |u_n|^p\rangle = T^{-1}\int_0^T|u_n|^pdt$, this yields the contribution of order 
\begin{equation}
T^{-1}v_n^p \Delta t_n = (Tk_n)^{-1}v_n^{p-1},
\label{eqM2b}
\end{equation} 
leading naturally to Eq.~(\ref{eqM2}). Hence, the functions $S'_p$ are expected to scale in the same way  as $S_p$ in the inertial range, i.e.,
\begin{equation}
S'_p(k_n) \propto k_n^{-\zeta_p}
\label{eqM3}
\end{equation}
with the same scaling exponents as in Eq.~(\ref{eqM}). This hypothesis perfectly agrees with the numerical simulations as shown in Figs.~\ref{fig3} and \ref{fig4}.

\begin{figure}
\centering \includegraphics[width = 0.6\textwidth]{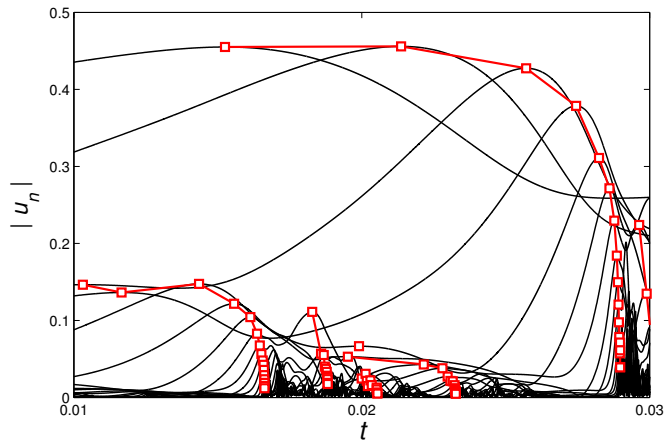}
\caption{(Color online) Typical dynamics of speed amplitudes $|u_n(t)|$ shown for the shells $n = 7,\ldots,24$. Red squares mark correlated sequences of local maxima (instantons), which have the structure similar to the blowup in Fig.~\ref{fig1}. Shown are the instantons created in shells $n_0 = 7,\ldots,14$.}
\label{fig5}
\end{figure}

The blowup in the inviscid shell model can be identified as the correlated sequence of maxima, which follow in increasing order of $n$ and $t$, see Fig.~\ref{fig1}. Analogous correlated structures (called the instantons) are observed in the inertial range of shell models~\cite{daumont2000instanton,gilson1997towards,l2001outliers}, see Fig.~\ref{fig5}. Following~\cite{mailybaev2012computation}, we identify the instanton as a sequence of local maxima $v_n = \max_t |u_n(t)|$ 
at times $t_n$ following in increasing order $t_{n_0} \le t_{n_0+1} \le \cdots \le t_{n_1}$. In this definition, no maxima of $|u_{n}(t)|$ or $|u_{n+1}(t)|$ are allowed in the interval $t_{n} < t < t_{n+1}$. Each instanton is created at some shell number $n_0$ and either reaches the viscous range or annihilates at a shell number $n_1$ in the inertial range. Using this rule, we group all maxima of velocity amplitudes into instantons, Fig.~\ref{fig5}. 

\begin{figure}
\centering \includegraphics[width = 0.7\textwidth]{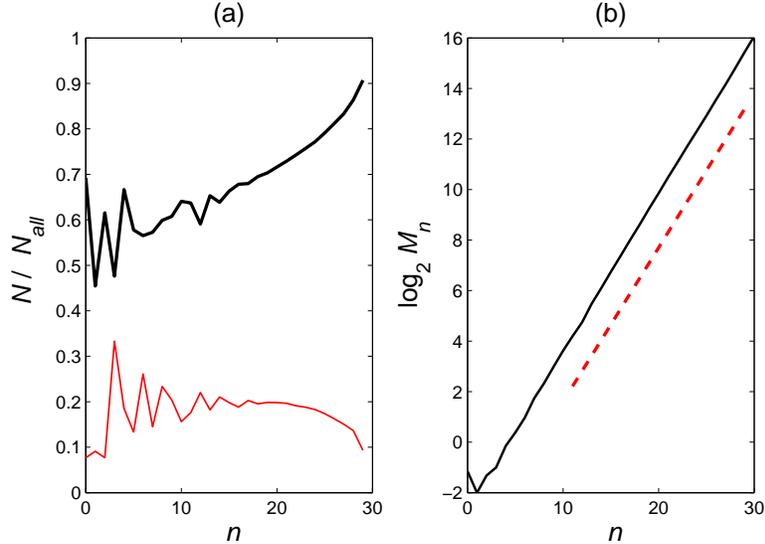}
\caption{(Color online) (a) The number $N$ of selected local maxima in the shell $n$ relative to their total number $N_{all}$. The bold black line corresponds to the maxima chosen from stable instantons. 
The thin red (lower) line corresponds to uncorrelated maxima (instantons of length 1 or 2). 
(b) Power-law scaling for the number $M_n$ of stable instantons created in shell $n$ per unit time. The slope $1-\zeta_1$ is shown by the dotted line.} 
\label{fig6}
\end{figure}

As we already mentioned, an instanton can be viewed as a blowup deformed by the inertial 
range environment, in which it propagates. One can see from Fig.~\ref{fig5} that this deformation 
is caused, mostly, by interaction with adjacent instantons. Let $N_{all}$ be the number of all maxima $v_n$ in a given shell $n$. 
Figure~\ref{fig6}a 
provides numerical values for the relative number $N/N_{all}$, where $N$ is the number of maxima in shell $n$ corresponding to a specified type of instantons. Most of the maxima correspond to stable instantons, 
which reach the viscous range, i.e., in our simulation $n_1 \ge 32$. These instantons cover from 60 to 90\% of the total number of maxima in a given shell $n$ (bold black line in Fig.~\ref{fig6}a). 
Majority of the remaining maxima (about 20\%) belong to very short instantons with $n_1 \approx n_0$, 
which can be considered as uncorrelated fluctuations. 
The instantons annihilating after traversing more than 2 shells but before the viscous range are rare.  

We see that the turbulent dynamics 
in inertial range of the Sabra model has the highly correlated structure, where the blowup plays a role of the driving mechanism. Another evidence supporting our observation is obtained if we compute 
the scaling exponents $\zeta_p$ for the functions (\ref{eqM2}), 
where only the maxima from stable instantons are included in the sum. 
These results are shown by crosses in Figs.~\ref{fig3} and \ref{fig4}. 
The same scaling exponents as for the velocity moments (\ref{eqM}) are obtained 
(a tiny difference in Fig.~\ref{fig4} is the same for all $p$ and, thus, 
corresponds to a small change of the total number of maxima included in the sum).   
In the following analysis we will consider only the maxima belonging to stable 
instantons in the sum~(\ref{eqM2}).

Description of the inertial range in terms of instantons provides a new interpretations 
of the first scaling exponent $\zeta_1$. Let $M_n$ be the average number of stable instantons 
created in shell $n$ per unit time. 
Using Eqs.~(\ref{eqM2}) and (\ref{eqM3}) we have  
\begin{equation}
S'_1(k_n) = \frac{1}{Tk_n}\sum 1 
= \frac{1}{k_n}\sum_{m = 0}^n M_m \propto k_n^{-\zeta_1},
\label{eqM2c}
\end{equation} 
where the first sum counts the maxima of stable instantons in shell $n$.
It is easy to check that Eq.~(\ref{eqM2c}) implies
\begin{equation}
M_n \propto k_n^{1-\zeta_1}.
\label{eqM3b}
\end{equation} 
We found that $\zeta_1$ determines the power-law scaling for the number of instantons 
created in shell $n$. For the Sabra model with $c = -0.5$, 
we have $1-\zeta_1 = 0.61$ in very good agreement with numerical data, Fig.~\ref{fig6}b.

The scaling exponent $\zeta_0 = 0$ is a simple consequence of the equality $\langle|u_n|^0\rangle = 1$. However, this exponent gets nontrivial interpretation in terms of velocity maxima in Eq.~(\ref{eqM2}) written as
\begin{equation}
S'_0(k_n) = \frac{1}{Tk_n}\sum v_n^{-1} \sim \frac{1}{T}\sum \Delta t_n,
\label{eqM4}
\end{equation} 
where, as we showed earlier, $\Delta t_n \sim (k_nv_n)^{-1}$ is the characteristic time associated with the maximum $v_n$. Relation $S'_0 \propto k_n^{\zeta_0} = 1$ implies that the total fraction of time occupied by these maxima is finite for each shell, i.e., the stable instantons are dense in space-time.

Numerical simulations for the model with the parameter 
values $c = -0.4$ and $-0.6$ were also carried out. The results are very similar to those 
presented in Figs.~\ref{fig3}--\ref{fig6}, which confirm our conclusion about the role of instantons 
as a principal elements of turbulent dynamics in the inertial range of the Sabra model.

\section{Self-similar statistics of instantons}
\label{sec5}

\begin{figure}
\centering \includegraphics[width = 0.7\textwidth]{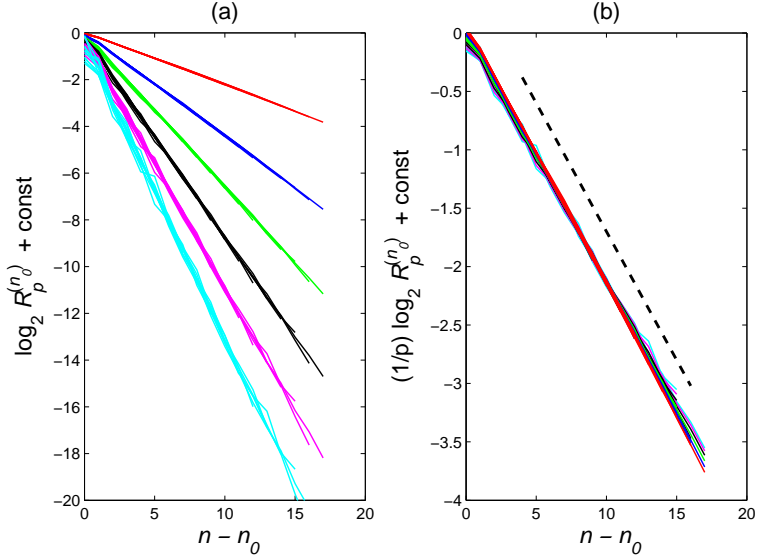}
\caption{(Color online) (a) The functions $R_p^{(n_0)}(k_n)$ in logarithmic coordinates demonstrating 
power-law scaling of instantons. Curves of the same color correspond to the instantons created in shells $n_0 = 13,\ldots,23$. Different colors indicate different values of $p = 1,\ldots,6$ from top to bottom. 
(b) Graphs of the left figure 
collapse onto a single straight line when divided by $p$. The slope $-y$ is shown by the dotted line.}
\label{fig7}
\end{figure}

The universal self-similarity of blowup (\ref{eq21}) is destroyed in the turbulent regime due to chaotic emergence and interaction of instantons, Fig.~\ref{fig5}. The most striking property of the instantons is that they restore the blowup self-similarity in statistical sense in the inertial range.
To observe this property, let us consider the functions
\begin{equation}
R_p^{(n_0)}(k_n) =
\frac{1}{T}\sum_{(n_0)} v_n^{p}, 
\quad n \ge n_0,
\label{eq9}
\end{equation}
where the sum is taken over the local maxima belonging to stable instantons created in fixed shell $n_0$. These functions can be viewed as effective velocity moments for the instantons born in a specific shell, and their graphs obtained numerically 
are shown in Fig.~\ref{fig7}a in logarithmic coordinates. One can clearly see that the functions $R_p^{(n_0)}$ obey the power-law scaling with exponents (slopes) independent of the initial shell number $n_0$.   

The next observation is that the slopes of the graphs in Fig.~\ref{fig7}a are proportional to $p$. This is shown in Fig.~\ref{fig7}b, where the functions $(1/p)\log_2 R_p^{(n_0)}$ are plotted versus a number of shells $n-n_0$ traversed by the instanton. All curves (after the vertical shift) 
collapse onto a single straight line of slope $-y$ with $y \approx 0.22$. This implies the relation 
\begin{equation}
R_p^{(n_0)}(k_n) = c_p^{(n_0)}\lambda^{-py\Delta n}
\propto k_{\Delta n}^{-py},\quad \Delta n = n-n_0 \ge 0,
\label{eq10}
\end{equation}
with the universal value of scaling exponent $y$ in the inertial range. The scaling exponent $y \approx 0.22$ is different but close to the scaling exponent $y_0 \approx 0.281$ of the blowup, see Fig.~\ref{fig2}a.  

\begin{figure}
\centering \includegraphics[width = 0.5\textwidth]{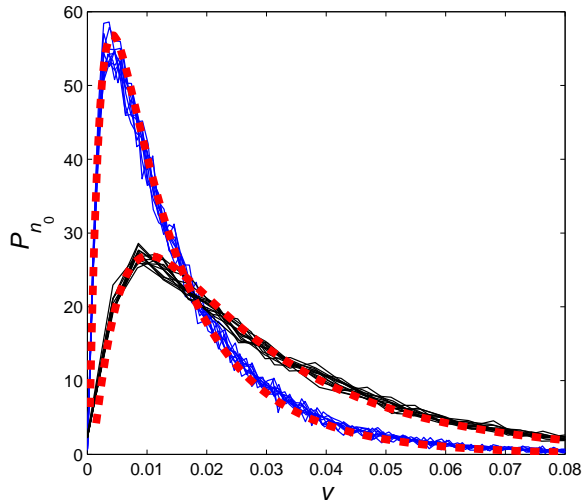}
\caption{(Color online) Renormalized PDFs $P_{n_0}(v)$ of instantons found numerically for $n_0 = 17$ and $n = 17,\ldots,29$ 
(thin black curves) and for $n_0 = 20$ and $n = 20,\ldots,29$ (thin blue curves). 
Collapse of the graphs with fixed $n_0$ onto a single curve confirms the self-similarity of PDFs in inertial range. 
Bold dotted curves show the PDFs determined by the large deviation principle.}
\label{fig8}
\end{figure}

The scaling rule in Eq.~(\ref{eq10}) suggests the universal self-similarity of instanton statistics. Let us consider the probability density functions (PDFs) determining the probability $P_{n_0,n}(v)dv$ to sample a local maximum $v = \max_t|u_n(t)|$ belonging to the instanton created in shell $n_0$. 
The self-similarity for PDFs implies that the renormalized function
\begin{equation}
P_{n_0}(v) = \lambda^{-y\Delta n}P_{n_0,n}(\lambda^{-y\Delta n}v)
\label{eq10b}
\end{equation}
does not depend on $n$ in the inertial range. This hypothesis fully agrees with the numerical results as one can see in Fig.~\ref{fig8}a, where the functions (\ref{eq10b}) for different $n$ 
collapse onto a single curve for fixed $n_0 = 17$ or $20$. The functions $P_{n_0}(v)$ for different $n_0$ are related by the large deviation principle, as we will show in the next section. 

We conclude that the instantons created in a given shell possess self-similar statistics. These instantons can be viewed as the blowup phenomena, which propagate to the viscous range interacting with each other. Interaction is an important factor which leads to a small but finite difference  between the scaling exponent of the instanton $y \approx 0.22$ and the scaling exponent of the blowup $y_0 \approx 0.28$. Similar results are obtained for the Sabra model with the parameters $c = -0.4$ and $-0.6$. The corresponding values of scaling exponents $y$ are shown in Fig.~\ref{fig2}a.

\section{Large deviation principle for instanton distributions}
\label{sec6}

According to Eqs.~(\ref{eqM2}), (\ref{eq9}) and (\ref{eq10}), an average contribution of a single instanton to the function $S'_p(k_n)$ is proportional to $k_n^{-1-(p-1)y}$. This yields an upper bound for the scaling exponents in Eq.~(\ref{eqM3}) as
\begin{equation}
\zeta_p \le 1+(p-1)y.
\label{eq20}
\end{equation}
The dotted line in Fig.~\ref{fig4} represents the right-hand side of Eq.~(\ref{eq20}). 
Since the graph of $\zeta_p$ is a concave function~\cite{frisch1995turbulence}, 
we conclude that the instanton scaling exponent $y$ does not determine any part of the $\zeta_p$ graph. 
In particular, $y \approx 0.22$ is larger than the slope of the $\zeta_p$ graph 
for large $p$ (the numerical data provides the slope $d\zeta_p/dp$ decreasing below 0.19). 
Therefore, the instanton scaling does not determine the scaling of high-order velocity moments,  
as it was conjectured in~\cite{l2001outliers} (however, this becomes true for different values of the model parameter $c$, as we show 
in Section~\ref{sec8}).  

The anomalous exponents $\zeta_p$ arise in the process of instanton creation. In order to see this, we use relations (\ref{eqM2}), (\ref{eq9}) (\ref{eq10}) and find
\begin{equation}
\begin{array}{rl}
S'_{p}(k_n) & 
\displaystyle
= k_n^{-1}\sum_{n_0 = 0}^n R_{p-1}^{(n_0)}(k_n) 
\\&
\displaystyle
= k_n^{-1}\sum_{n_0 = 0}^n c_{p-1}^{(n_0)}\lambda^{-(p-1)y(n-n_0)}.
\end{array}
\label{eq11}
\end{equation}
Then the coefficients are expressed from 
(\ref{eq11}) as
\begin{equation}
c_{p-1}^{(n)} 
= k_nS'_p(k_n)-\lambda^{-(p-1)y}k_{n-1}S'_p(k_{n-1}).
\label{eq13}
\end{equation}
In the inertial range, where the power-law scaling (\ref{eqM3}) holds, we have
\begin{equation}
c_{p-1}^{(n)} \propto k_n^{1-\zeta_p}.
\label{eq12}
\end{equation}
This relation was also confirmed numerically. We see that, due to the self-similar structure of instantons, anomalous scaling is attributed 
exclusively to the coefficients $c_p^{(n_0)}$ describing amplitudes of instantons created in shell $n_0$. This property relates the inertial range anomaly with the process of instanton creation. 

Relation (\ref{eq12}) allows finding the universal form of PDFs $P_n(v)$ in Eq.~(\ref{eq10b}) as follows. Using Eqs.~(\ref{eq9})--(\ref{eq10b}), we obtain \begin{equation}
\begin{array}{rcl}
c_{p-1}^{(n)} 
& = & \displaystyle
R_{p-1}^{(n)}(k_n) 
= \frac{1}{T}\sum_{(n)} v_n^{p-1} \\[17pt]
& = & \displaystyle
M_n\int_0^\infty v^{p-1}P_{n,n}(v) dv
= M_n\int_0^\infty v^{p-1}P_n(v) dv,
\end{array}
\label{eq15}
\end{equation}
where $M_n$ is the number of instantons created in shell $n$ per unit time.
Introducing the new variable $a$ and function $\rho(a)$ as 
\begin{equation}
a = \frac{1}{n}\log_\lambda \frac{v}{v_*},\quad 
\rho_n(a) = \rho_* nM_nP_n(v),
\label{eq14}
\end{equation}
where $v_*$ and $\rho_*$ are constant coefficients,
we write Eq.~(\ref{eq15}) in the form

\begin{equation}
c_{p-1}^{(n)} = \frac{v_*^p}{\rho_*}\log\lambda \int \lambda^{npa}\rho_n(a) da. 
\label{eq16}
\end{equation}
Using Eq.~(\ref{eq12}), we find the power-law scaling rule for the integral in the right-hand side as 
\begin{equation}
\int \lambda^{npa}\rho_n(a) da \propto 
k_n^{1-\zeta_p} = \lambda^{n(1-\zeta_p)}.
\label{eq18}
\end{equation}

In Eq.~(\ref{eq18}) the scaling exponent $1-\zeta_p$ is a smooth convex function of $p \in \mathbb{R}$, Fig.~\ref{fig4}.
Hence, we can apply the G\"artner-Ellis theorem~\cite{gartner1977large,ellis1984large,touchette2009large} to  Eq.~(\ref{eq18}), which states that $\rho_n(a)$ has the asymptotic form 
\begin{equation}
\rho_n(a) \propto \lambda^{-nJ(a)} = k_n^{-J(a)}
\label{eq23}
\end{equation}
for large $n$, where the rate function $J(a)$ is the Legendre transform of the function $1-\zeta_p$, i.e.,
\begin{equation}
J(a) = pa-(1-\zeta_p), \quad a = -\frac{d\zeta_p}{dp}.
\label{eq24}
\end{equation}
Expression (\ref{eq23}) is called the large deviation principle.
Note that the G\"artner-Ellis theorem is formulated for $\lambda = e$ but one can easily check its validity for any $\lambda > 1$. 

\begin{figure}
\centering \includegraphics[width = 0.5\textwidth]{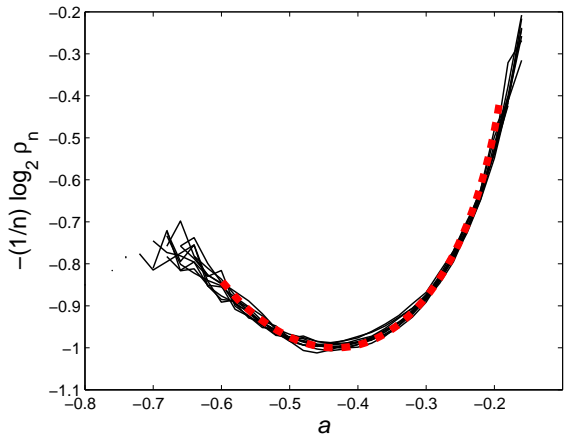}
\caption{(Color online) The functions $-(1/n)\log_\lambda\rho_n$ computed numerically for $n = 17,\ldots,26$ (thin black lines) are compared with the rate function $J(a)$ 
(dotted red line).}
\label{fig9}
\end{figure}

We verify Eq.~(\ref{eq23}) in Fig.~\ref{fig9}, where the black curves show the functions $-(1/n)\log_\lambda \rho_n(a)$ found numerically using Eqs.~(\ref{eq10b}) and (\ref{eq14}) for $n =  17,\ldots,26$ and $n_0 = n$. As expected, these graphs collapse 
onto a single curve given by the rate function $J(a)$. 
The rate function represented by the red dotted line was computed using the Legendre transform (\ref{eq24}) for the scaling exponent $\zeta_p$ in the interval $-2 \le p \le 10$. In numerical computations, it was important to choose good values of the constants $v_*$ and $\rho_*$ in Eq.~(\ref{eq14}) in order to achieve better convergence.

The final result of our derivation is obtained by substituting Eq.~(\ref{eq23}) 
into (\ref{eq14}) as
\begin{equation}
M_nP_n(v) \propto k_n^{-J(a)},\quad 
a = \frac{1}{n}\log_\lambda \frac{v}{v_*},
\label{eq25}
\end{equation} 
where we dropped the factor $n^{-1}$ representing a logarithmic correction for the first expression.
Note that the asymptotic form given by the G\"artner-Ellis theorem in Eq.~(\ref{eq23}) is understood as $n^{-1}\log_\lambda\rho_n(a) \to -J(a)$ in the limit $n \to \infty$. Recall that the same limit of large $n$ is used in the definition of inertial range, which corresponds to shell numbers far from the forcing range. Thus, Eq.~(\ref{eq25}) is valid in the inertial range.
This statement is confirmed numerically in Fig.~\ref{fig8}, where the asymptotic PDFs given by Eq.~(\ref{eq25}) are shown by the dotted red curves for the shells $17$ and $20$ (with the constant factors properly adjusted). 

We showed that the PDFs of instantons in the inertial range have the universal self-similar form (\ref{eq25}) related to the anomalous scaling exponents by Eqs.~(\ref{eq24}). Thus, the instantons satisfy the large deviation principle leading to the inertial range anomaly. 
The presented analysis has much in common with the phenomenological model of multifractality~\cite{frisch1995turbulence}. In this model, it is assumed that the velocity field can be decomposed into fractal subsets with different scaling properties, and the fractal dimensions are related to the anomalous exponents $\zeta_p$ by the Legendre transform. However, the fractal subsets in the multifractal model are hard to define and detect numerically or experimentally, as well as to justify their appearance. On the contrary, the presented approach based on the study of instantons is related to the analytical theory of blowup and is supported by the detailed numerical analysis. 

\section{Instanton creation model}
\label{sec7}

In this section, we introduce a phenomenological model for instanton creation, 
where the large deviation principle can be derived analytically. 
As one can see in Fig.~\ref{fig5}, an instanton traveling through the inertial range leaves a trace (energy) in all the shells it passed through. 
Due to the asymptotic stability of blowup mentioned in Section~\ref{sec3}, 
this energy ``feeds'' a series of newly created instantons in different shells. 
This process leads to formation of a ``gas'' of instantons, 
which is dense in space-time and carries the energy from the forcing to the viscous range. 
As the viscosity plays no role in this process, the dissipation anomaly becomes a natural consequence of the described behavior.

In this phenomenological picture, instantons create other instantons.
A simple statistical model of the creation process can be developed as follows. 
We assume that an instanton, which reaches the shell $n$ with the amplitude $v_n = 1$, 
creates in average $\varphi(v)dv$ new instantons of amplitude $v_n = v$ in this shell. 
Here $\varphi(v)$ is the creation rate function, which is assumed to be universal, i.e., independent of $n$. 
For an instanton of arbitrary amplitude $v_n = v'$, the density of created instantons is given by
\begin{equation}
\frac{1}{v'}\,\varphi\left(\frac{v}{v'}\right)dv,
\label{eq30}
\end{equation}
as it follows from the scaling symmetry of the Sabra model.
As before, we consider only stable instantons, which cover up to $90$\% of all 
shell oscillations (Fig.~\ref{fig6}a), and disregard other types of fluctuations. 

Using the definitions of Section~\ref{sec5}, the distribution of instanton amplitudes is described by the product 
\begin{equation}
M_nP_{n,n}(v)dv,
\label{eq31}
\end{equation}
determining a number of instantons with maxima $v_n = v$ created in shell $n$ per unit time. 
Here $P_{n,n}$ is the PDF of such instantons and $M_n$ is the total instanton creation rate in shell $n$. 
Distribution of maxima $v_n = v$ corresponding to the instantons created in previous shells $n_0 < n$ is found similarly as
\begin{equation}
\sum_{n_0 = 0}^{n-1} M_{n_0}P_{n_0,n}(v)dv.
\label{eq32}
\end{equation}
Using expressions (\ref{eq31}) and (\ref{eq32}), the instanton creation principle described by Eq.~(\ref{eq30}) yields
\begin{equation}
M_nP_{n,n}(v) = \int_0^\infty 
\left[\sum_{n_0 = 0}^{n-1} M_{n_0}P_{n_0,n}(v')\right]
\frac{1}{v'}\,\varphi\left(\frac{v}{v'}\right)dv'.
\label{eq33}
\end{equation}
Using Eq.~(\ref{eq10b}) we write this expression as
\begin{equation}
M_nP_n(v) 
= \sum_{n_0 = 0}^{n-1} \int_0^\infty 
\lambda^{y(n-n_0)}M_{n_0}P_{n_0}(\lambda^{y(n-n_0)}v')
\frac{1}{v'}\,\varphi\left(\frac{v}{v'}\right)dv'.
\label{eq34}
\end{equation}

Let us compute the quantity (\ref{eq15}) represented as
\begin{equation}
c_p^{(n)} = \int_0^\infty v^p M_nP_n(v)dv.
\label{eq35}
\end{equation}
Expressing $M_nP_n$ from Eq.~(\ref{eq34}) and 
denoting $\xi = v/v'$ and $\eta = \lambda^{y(n-n_0)}v'$, 
we find
\begin{equation}
c_p^{(n)} 
= \sum_{n_0 = 0}^{n-1} 
\lambda^{-yp(n-n_0)}
\iint_0^\infty (\eta\xi)^p M_{n_0}P_{n_0}(\eta)
\varphi(\xi)\,d\xi\,d\eta.
\label{eq36}
\end{equation}
Using Eq.~(\ref{eq35}) for $c_p^{(n_0)}$,
the right-hand side of Eq.~(\ref{eq36}) is integrated as
\begin{equation}
c_p^{(n)}
= \varphi_p\sum_{n_0 = 0}^{n-1} \lambda^{-yp(n-n_0)}
c_p^{(n_0)}, 
\label{eq37}
\end{equation}
where
\begin{equation}
\varphi_p = \int_0^\infty \xi^p\varphi(\xi)d\xi.
\label{eq37b}
\end{equation}
Using Eq.~(\ref{eq37}) for $c_p^{(n)}$ and $c_p^{(n-1)}$, we compute the following difference
\begin{equation}
c_p^{(n)}-\lambda^{-yp}c_p^{(n-1)}
= \varphi_p\lambda^{-yp}c_p^{(n-1)}, 
\label{eq38}
\end{equation}
and obtain
\begin{equation}
c_p^{(n)} 
= \lambda^{-yp}(1+\varphi_p)c_p^{(n-1)}. 
\label{eq39}
\end{equation}
Expression (\ref{eq39}) implies the power-law (\ref{eq12}) with the exponents
\begin{equation}
\zeta_p = 1+(p-1)y-\log_\lambda(1+\varphi_{p-1}).
\label{eq40}
\end{equation}
Since $\varphi_{p-1} > 0$, this expression satisfies the inequality (\ref{eq20}). 

We determined 
the scaling exponents $\zeta_p$ explicitly in terms of the moments of 
the creation rate function $\varphi(v)$. For a particular example, when the creation rate is given by self-similarity arguments, this computation was done in~\cite{mailybaev2012computation}.  
As we showed in Section~\ref{sec6}, if $\zeta_p$ exist 
and are differentiable for all real values of $p$, then the G\"artner-Ellis theorem ensures 
the large deviation principle (\ref{eq25}). Therefore, the large deviation principle in our 
model is verified for any creation rate function, which has finite and differentiable moments $\varphi_p$ for all real $p$.

We proved that the large deviation principle emerges naturally for a class of simple models of instanton creation.
These models have several essential simplifications. In particular, we disregarded correlations in time and correlations between instantons in different shells.
Numerical simulations show that such correlations are important. In fact, numerical values for the moments $\varphi_p$ deviate strongly from the values determined by Eq.~(\ref{eq40}) for known $\zeta_p$. Also, the function $\varphi(v)$ depends on a way it is computed. On the other hand, the numerical data provided good evidence for universality of the creation process, because the function $\varphi(v)$ found numerically does not depend on the shell number $n$. 

\section{Turbulent regime dominated by a single blowup}
\label{sec8}

We showed in Section~\ref{sec5} that the exponent $y$ describing the universal scaling of instantons in Eq.~(\ref{eq10b}) does not determine any of the scaling exponents $\zeta_p$, neither their asymptotic behavior for large $p$, see Eq.~(\ref{eq20}) and Fig.~\ref{fig4}. The anomalous scaling of velocity moments 
is linked exclusively to the process of instanton creation. 
In this section we demonstrate a different turbulent regime, where the instanton scaling has strong influence 
on the velocity moments. This regime can be predicted by looking at the blowup scaling exponent $y_0$ depending 
on the model parameter in Fig.~\ref{fig2}a. At $c = -0.139$, we have $y_0 = 0$. 
Hence, in the asymptotic form of blowup given by Eq.~(\ref{eq21}), 
all the local maxima $v_n = \max_t|u_n(t)|$ are of the same order of magnitude. 
As a result, a single blowup provides the terms $(Tk_n)^{-1}v_n^{p-1} \sim k_n^{-1}$ in the 
sum (\ref{eqM2}) for the structure functions $S'_p$, which  
yields the condition $\zeta_p \le 1$ for all $p$. This upper bound is exact for  
$\zeta_3 = 1$, which is required by the existence of energy cascade. 
This fact suggests that the blowup scaling should play essential role for the Sabra models 
with the parameter $c$ in the neighborhood of $-0.139$. 

\begin{figure}
\centering \includegraphics[width = 0.5\textwidth]{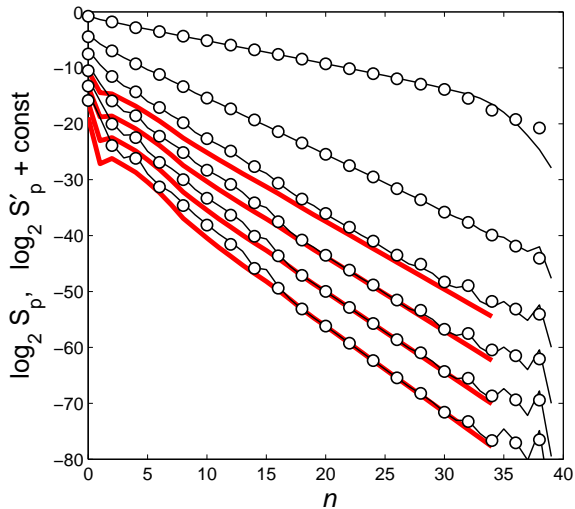}
\caption{(Color online) Thin black curves present the velocity moments $S_p(k_n) = \langle |u_n|^p\rangle$ 
for $p = 1,3,5,7,9,11$. Circles determine the functions $S'_p(k_n)$ 
from Eq.~(\ref{eqM2}) for the same $p$ and even $n$ (vertical shifts are applied to facilitate comparison with $S_p$).
The red (bold gray) curves show the values $(Tk_n)^{-1}v_n^{p-1}$ for a single dominant instanton and $p = 5,7,9,11$.}
\label{fig10}
\end{figure}

Figure~\ref{fig10} (thin lines and circles) shows the structure functions $S_p(k_n)$ and $S'_p(k_n)$ for $c = -0.2$ 
in logarithmic coordinates. 
Vertical shifts are used to compare the graphs for $S_p(k_n)$ and $S'_p(k_n)$, 
and the good match confirms validity of the description based on velocity maxima.
The inertial range 
shrinks substantially 
in this model and corresponds roughly to the shells $17 \le n \le 30$.
In this case the blowup scaling exponent $y_0 = 0.0534$ is small. 
As we just mentioned, due to a slow decay of blowup amplitudes in Eq.~(\ref{eq21}), we expect that the inertial 
range is influenced by the blowup scaling. 

The numerical simulation shows that a single stable instanton dominates the inertial range.  
This instanton is created in the initial shell $n_0 = 0$ and travels all the way to the viscous range. 
The bold red (gray) curves in Fig.~\ref{fig10} show the values of a specific term $(Tk_n)^{-1}v_n^{p-1}$ 
in the sum (\ref{eqM2}), which corresponds to this instanton. 
In the figure, we used the same vertical shift for the red curve 
as for the full sum $S'_p(k_n)$, which shows that not only the slope but also the value of $S'_p(k_n)$ 
is determined by a single instanton for large $p$. Due to its dominant role, 
this instanton is weakly influenced by surrounding 
fluctuations, i.e., by other instantons. As a result, we can expect that the dominant instanton scales with the same exponent $y_0$
as the blowup. 
This hypothesis agrees perfectly with the numerical data.

\begin{figure}
\centering \includegraphics[width = 0.5\textwidth]{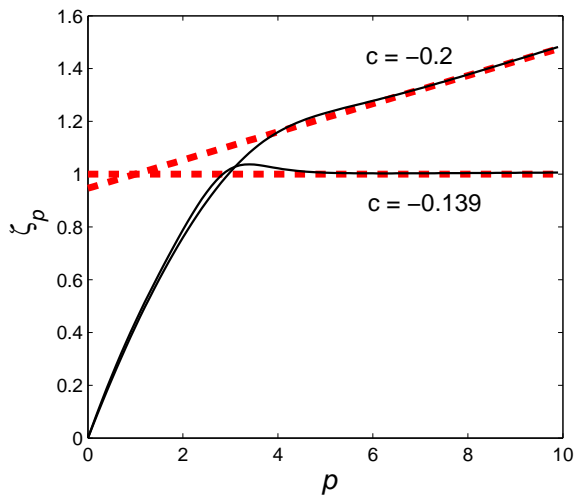}
\caption{(Color online) Anomalous scaling exponent $\zeta_p$ computed for the Sabra model with $c = -0.2$ 
and $-0.139$. For each case, the red dotted line shows the values of $\zeta_p$ determined 
by the universal scaling of a single blowup.}
\label{fig11}
\end{figure}

Figure~\ref{fig11} shows the scaling exponents $\zeta_p$ 
for the Sabra model with $c = -0.2$ and $-0.139$. In this case 
the blowup scaling exponent is equal to $y_0 = 0.0534$ and $0$, respectively, see also Fig.~\ref{fig2}a. The dotted straight lines in Fig.~\ref{fig11} show the right-hand side of the inequality (\ref{eq20}) with $y = y_0$. We see that this inequality becomes exact for large $p$ because of the dominant role of 
a single instanton. Note that the graph of $\zeta_p$ is not concave, 
as it must be, and violates slightly Eq.~(\ref{eq20}) in the region near the intersection with the dotted line. This seems to be a numerical artifact due to very slow convergence 
in the region where the blowup scaling competes with the scaling of instanton creation process. The horizontal part of the graph with $\zeta_p = 1$ for $c = -0.139$ reminds the analogous behavior of scaling exponents for turbulence of the Burgers equation, see~\cite{aurell1992multifractal,cardy2008non}. 
Our results show that there are parameter values 
of the Sabra model, when the anomalous scaling exponents $\zeta_p$ are 
explicitly related to the blowup scaling exponent $y_0$ for some range of $p$. 

\section{Conclusion}
\label{sec9}

In this paper we have shown that the blowup (a singularity developing in finite time) 
may be considered as a basic element in the theory of developed turbulence 
for the Sabra shell model. We utilize the fact that the blowup in the inviscid system 
has universal asymptotic form, where shell speeds pass successively through their maxima 
with increasing time and wavenumber. 
This sequence of maxima is the main property used for identifying and analyzing  
the coherent turbulent bursts, which are induced by blowup and called the instantons.

Blowup is characterized by the asymptotically stable traveling wave solution for the renormalized system. Thus,
almost any initial condition of finite norm leads to blowup in the inviscid Sabra model. 
In the inertial range of turbulent regime, the blowup-like structures appear in every available part 
of space-time and propagate toward the viscous range. This dynamical behavior can be viewed 
as a ``gas'' of instantons, which is dense in space-time and moves from the forcing to the viscous range. 
Existence of many interacting instantons alter their properties, as compared to the ``pure'' blowup, but 
the instanton statistics remains self-similar and universal.

We showed that the anomalous scaling of velocity moments is a natural consequence of the 
instanton creation process, which obeys the large deviation principle. This allows, 
in particular, justifying the universal form 
of instanton probability density functions in inertial range and describing these universal functions analytically 
in terms of the scaling exponents of velocity moments. The obtained results are in excellent agreement with the numerical data.

The described dynamical picture brings us back to the famous Richardson description~\cite{richardson2007weather}: 
``Big whirls have little whirls that feed on their velocity, 
and little whirls have lesser whirls and so on to viscosity'', which is known to be inadequate for the 
Navier-Stokes turbulence~\cite{frisch1995turbulence}. We see now that, in the Sabra shell model, 
this description becomes true if one substitutes ``whirls'' by ``instantons'' (or ``blowups''). 
The dissipation anomaly follows naturally from this picture, because the instanton dynamics 
is unrelated to viscosity in the inertial range and the instantons 
move only toward large wavenumbers. The intermittency becomes a simple consequence 
of the instanton scaling, where the scales of time and velocity are related 
as $\Delta t_n \sim (k_nv_n)^{-1}$. This means that large-amplitude instantons are fast, while 
small-amplitude instantons are slow and can be viewed as windows of low activity.
We showed, however, that the described scenario 
is not the only possibility, and the anomalous exponents may be linked to universal properties 
of a single blowup for specific values of the model parameter.

The essential part of our work is based on numerical data. Here 
the blowup, whose properties follow from the model equations, is 
used as a guideline for the numerically accessible definition of instantons. 
If the analytical theory of turbulence for the shell model can be developed in a similar manner, 
the formal definition of instantons and derivation of their universal scaling properties 
directly from the shell model equations would be the major difficulty. 
An important step in this direction was done in~\cite{daumont2000instanton}, where self-similar statistics of 
instantons was derived as a result of the blowup interacting with small random fluctuations. 
Our results show that this theory should be extended by taking into account interactions among instantons and instanton creation. Note that, for different 
shell models, the blowup structure may be not self-similar 
but still universal~\cite{mailybaev2012c} providing an extra complication. 
It would be interesting to test these ideas also on the cascade models of turbulence, 
where the interactions among variables within the same shell are introduced~\cite{uhlig1997local,uhlig1997singularities}.

As for the 3D Navier-Stokes developed turbulence, our results confirm the common understanding of 
the importance of coherent structures like, e.g., formation of vortex filaments. 
The novel idea is that it is the universal creation process of these structures what may drive 
turbulent dynamics in the inertial range, while the scaling of an individual structure plays a secondary role. 
Moreover, one may notice that such structures do not have to blow up in finite time 
in the inviscid limit and, e.g., the exponential rate would be sufficient. 
The method for identifying and tracking coherent structures from the moment of their creation till the viscous range has to be developed in order to verify our hypotheses numerically or experimentally.    
 

\bibliographystyle{apsrev4-1}
\bibliography{refs}

\begin{thebibliography}{30}%
\makeatletter
\providecommand \@ifxundefined [1]{%
 \@ifx{#1\undefined}
}%
\providecommand \@ifnum [1]{%
 \ifnum #1\expandafter \@firstoftwo
 \else \expandafter \@secondoftwo
 \fi
}%
\providecommand \@ifx [1]{%
 \ifx #1\expandafter \@firstoftwo
 \else \expandafter \@secondoftwo
 \fi
}%
\providecommand \natexlab [1]{#1}%
\providecommand \enquote  [1]{``#1''}%
\providecommand \bibnamefont  [1]{#1}%
\providecommand \bibfnamefont [1]{#1}%
\providecommand \citenamefont [1]{#1}%
\providecommand \href@noop [0]{\@secondoftwo}%
\providecommand \href [0]{\begingroup \@sanitize@url \@href}%
\providecommand \@href[1]{\@@startlink{#1}\@@href}%
\providecommand \@@href[1]{\endgroup#1\@@endlink}%
\providecommand \@sanitize@url [0]{\catcode `\\12\catcode `\$12\catcode
  `\&12\catcode `\#12\catcode `\^12\catcode `\_12\catcode `\%12\relax}%
\providecommand \@@startlink[1]{}%
\providecommand \@@endlink[0]{}%
\providecommand \url  [0]{\begingroup\@sanitize@url \@url }%
\providecommand \@url [1]{\endgroup\@href {#1}{\urlprefix }}%
\providecommand \urlprefix  [0]{URL }%
\providecommand \Eprint [0]{\href }%
\providecommand \doibase [0]{http://dx.doi.org/}%
\providecommand \selectlanguage [0]{\@gobble}%
\providecommand \bibinfo  [0]{\@secondoftwo}%
\providecommand \bibfield  [0]{\@secondoftwo}%
\providecommand \translation [1]{[#1]}%
\providecommand \BibitemOpen [0]{}%
\providecommand \bibitemStop [0]{}%
\providecommand \bibitemNoStop [0]{.\EOS\space}%
\providecommand \EOS [0]{\spacefactor3000\relax}%
\providecommand \BibitemShut  [1]{\csname bibitem#1\endcsname}%
\let\auto@bib@innerbib\@empty
\bibitem [{\citenamefont {Kolmogorov}(1941)}]{kolmogorov1941local}%
  \BibitemOpen
  \bibfield  {author} {\bibinfo {author} {\bibfnamefont {A.}~\bibnamefont
  {Kolmogorov}},\ }\href@noop {} {\bibfield  {journal} {\bibinfo  {journal}
  {Dokl. Akad. Nauk SSSR}\ }\textbf {\bibinfo {volume} {30}},\ \bibinfo {pages}
  {9} (\bibinfo {year} {1941})}\BibitemShut {NoStop}%
\bibitem [{\citenamefont {Frisch}(1995)}]{frisch1995turbulence}%
  \BibitemOpen
  \bibfield  {author} {\bibinfo {author} {\bibfnamefont {U.}~\bibnamefont
  {Frisch}},\ }\href@noop {} {\emph {\bibinfo {title} {{Turbulence: The Legacy
  of A.N. Kolmogorov}}}}\ (\bibinfo  {publisher} {Cambridge University Press},\
  \bibinfo {year} {1995})\BibitemShut {NoStop}%
\bibitem [{\citenamefont {Cardy}\ \emph {et~al.}(2008)\citenamefont {Cardy},
  \citenamefont {Falkovich},\ and\ \citenamefont {Gawedzki}}]{cardy2008non}%
  \BibitemOpen
  \bibfield  {author} {\bibinfo {author} {\bibfnamefont {J.}~\bibnamefont
  {Cardy}}, \bibinfo {author} {\bibfnamefont {G.}~\bibnamefont {Falkovich}}, \
  and\ \bibinfo {author} {\bibfnamefont {K.}~\bibnamefont {Gawedzki}},\
  }\href@noop {} {\emph {\bibinfo {title} {{Non-equilibrium Statistical
  Mechanics and Turbulence}}}}\ (\bibinfo  {publisher} {Cambridge University
  Press},\ \bibinfo {year} {2008})\BibitemShut {NoStop}%
\bibitem [{\citenamefont {Eyink}\ and\ \citenamefont
  {Sreenivasan}(2006)}]{eyink2006onsager}%
  \BibitemOpen
  \bibfield  {author} {\bibinfo {author} {\bibfnamefont {G.}~\bibnamefont
  {Eyink}}\ and\ \bibinfo {author} {\bibfnamefont {K.}~\bibnamefont
  {Sreenivasan}},\ }\href@noop {} {\bibfield  {journal} {\bibinfo  {journal}
  {Rev. Modern Phys.}\ }\textbf {\bibinfo {volume} {78}},\ \bibinfo {pages}
  {87} (\bibinfo {year} {2006})}\BibitemShut {NoStop}%
\bibitem [{\citenamefont {Eyink}(1994)}]{eyink1994energy}%
  \BibitemOpen
  \bibfield  {author} {\bibinfo {author} {\bibfnamefont {G.}~\bibnamefont
  {Eyink}},\ }\href@noop {} {\bibfield  {journal} {\bibinfo  {journal} {Physica
  D}\ }\textbf {\bibinfo {volume} {78}},\ \bibinfo {pages} {222} (\bibinfo
  {year} {1994})}\BibitemShut {NoStop}%
\bibitem [{\citenamefont {Constantin}\ \emph {et~al.}(1994)\citenamefont
  {Constantin}, \citenamefont {Weinan},\ and\ \citenamefont
  {Titi}}]{constantin1994onsager}%
  \BibitemOpen
  \bibfield  {author} {\bibinfo {author} {\bibfnamefont {P.}~\bibnamefont
  {Constantin}}, \bibinfo {author} {\bibfnamefont {E.}~\bibnamefont {Weinan}},
  \ and\ \bibinfo {author} {\bibfnamefont {E.}~\bibnamefont {Titi}},\
  }\href@noop {} {\bibfield  {journal} {\bibinfo  {journal} {Commun. Math.
  Phys.}\ }\textbf {\bibinfo {volume} {165}},\ \bibinfo {pages} {207} (\bibinfo
  {year} {1994})}\BibitemShut {NoStop}%
\bibitem [{\citenamefont {Aurell}\ \emph {et~al.}(1992)\citenamefont {Aurell},
  \citenamefont {Frisch}, \citenamefont {Lutsko},\ and\ \citenamefont
  {Vergassola}}]{aurell1992multifractal}%
  \BibitemOpen
  \bibfield  {author} {\bibinfo {author} {\bibfnamefont {E.}~\bibnamefont
  {Aurell}}, \bibinfo {author} {\bibfnamefont {U.}~\bibnamefont {Frisch}},
  \bibinfo {author} {\bibfnamefont {J.}~\bibnamefont {Lutsko}}, \ and\ \bibinfo
  {author} {\bibfnamefont {M.}~\bibnamefont {Vergassola}},\ }\href@noop {}
  {\bibfield  {journal} {\bibinfo  {journal} {J. Fluid Mech.}\ }\textbf
  {\bibinfo {volume} {238}},\ \bibinfo {pages} {467} (\bibinfo {year}
  {1992})}\BibitemShut {NoStop}%
\bibitem [{\citenamefont {Gibbon}(2008)}]{gibbon2008three}%
  \BibitemOpen
  \bibfield  {author} {\bibinfo {author} {\bibfnamefont {J.~D.}\ \bibnamefont
  {Gibbon}},\ }\href@noop {} {\bibfield  {journal} {\bibinfo  {journal}
  {Physica D}\ }\textbf {\bibinfo {volume} {237}},\ \bibinfo {pages} {1894}
  (\bibinfo {year} {2008})}\BibitemShut {NoStop}%
\bibitem [{\citenamefont {Gledzer}(1973)}]{gledzer1973system}%
  \BibitemOpen
  \bibfield  {author} {\bibinfo {author} {\bibfnamefont {E.~B.}\ \bibnamefont
  {Gledzer}},\ }\href@noop {} {\bibfield  {journal} {\bibinfo  {journal} {Sov.
  Phys. Doklady}\ }\textbf {\bibinfo {volume} {18}},\ \bibinfo {pages} {216}
  (\bibinfo {year} {1973})}\BibitemShut {NoStop}%
\bibitem [{\citenamefont {Ohkitani}\ and\ \citenamefont
  {Yamada}(1989)}]{ohkitani1989temporal}%
  \BibitemOpen
  \bibfield  {author} {\bibinfo {author} {\bibfnamefont {K.}~\bibnamefont
  {Ohkitani}}\ and\ \bibinfo {author} {\bibfnamefont {M.}~\bibnamefont
  {Yamada}},\ }\href@noop {} {\bibfield  {journal} {\bibinfo  {journal} {Prog.
  Theor. Phys.}\ }\textbf {\bibinfo {volume} {89}},\ \bibinfo {pages} {329}
  (\bibinfo {year} {1989})}\BibitemShut {NoStop}%
\bibitem [{\citenamefont {L'vov}\ \emph {et~al.}(1998)\citenamefont {L'vov},
  \citenamefont {Podivilov}, \citenamefont {Pomyalov}, \citenamefont
  {Procaccia},\ and\ \citenamefont {Vandembroucq}}]{l1998improved}%
  \BibitemOpen
  \bibfield  {author} {\bibinfo {author} {\bibfnamefont {V.~S.}\ \bibnamefont
  {L'vov}}, \bibinfo {author} {\bibfnamefont {E.}~\bibnamefont {Podivilov}},
  \bibinfo {author} {\bibfnamefont {A.}~\bibnamefont {Pomyalov}}, \bibinfo
  {author} {\bibfnamefont {I.}~\bibnamefont {Procaccia}}, \ and\ \bibinfo
  {author} {\bibfnamefont {D.}~\bibnamefont {Vandembroucq}},\ }\href@noop {}
  {\bibfield  {journal} {\bibinfo  {journal} {Phys. Rev. E}\ }\textbf {\bibinfo
  {volume} {58}},\ \bibinfo {pages} {1811} (\bibinfo {year}
  {1998})}\BibitemShut {NoStop}%
\bibitem [{\citenamefont {Biferale}(2003)}]{biferale2003shell}%
  \BibitemOpen
  \bibfield  {author} {\bibinfo {author} {\bibfnamefont {L.}~\bibnamefont
  {Biferale}},\ }\href@noop {} {\bibfield  {journal} {\bibinfo  {journal}
  {Annu. Rev. Fluid Mech.}\ }\textbf {\bibinfo {volume} {35}},\ \bibinfo
  {pages} {441} (\bibinfo {year} {2003})}\BibitemShut {NoStop}%
\bibitem [{\citenamefont {Constantin}\ \emph {et~al.}(2007)\citenamefont
  {Constantin}, \citenamefont {Levant},\ and\ \citenamefont
  {Titi}}]{constantin2007regularity}%
  \BibitemOpen
  \bibfield  {author} {\bibinfo {author} {\bibfnamefont {P.}~\bibnamefont
  {Constantin}}, \bibinfo {author} {\bibfnamefont {B.}~\bibnamefont {Levant}},
  \ and\ \bibinfo {author} {\bibfnamefont {E.~S.}\ \bibnamefont {Titi}},\
  }\href@noop {} {\bibfield  {journal} {\bibinfo  {journal} {Phys. Rev. E}\
  }\textbf {\bibinfo {volume} {75}},\ \bibinfo {pages} {016304} (\bibinfo
  {year} {2007})}\BibitemShut {NoStop}%
\bibitem [{\citenamefont {Dombre}\ and\ \citenamefont
  {Gilson}(1998)}]{dombre1998intermittency}%
  \BibitemOpen
  \bibfield  {author} {\bibinfo {author} {\bibfnamefont {T.}~\bibnamefont
  {Dombre}}\ and\ \bibinfo {author} {\bibfnamefont {J.~L.}\ \bibnamefont
  {Gilson}},\ }\href@noop {} {\bibfield  {journal} {\bibinfo  {journal}
  {Physica D}\ }\textbf {\bibinfo {volume} {111}},\ \bibinfo {pages} {265}
  (\bibinfo {year} {1998})}\BibitemShut {NoStop}%
\bibitem [{\citenamefont {Mailybaev}(2013)}]{mailybaev2012c}%
  \BibitemOpen
  \bibfield  {author} {\bibinfo {author} {\bibfnamefont {A.~A.}\ \bibnamefont
  {Mailybaev}},\ }\href@noop {} {\bibfield  {journal} {\bibinfo  {journal}
  {Nonlinearity}\ }\textbf {\bibinfo {volume} {26}},\ \bibinfo {pages} {1105}
  (\bibinfo {year} {2013})}\BibitemShut {NoStop}%
\bibitem [{\citenamefont {Mailybaev}(2012{\natexlab{a}})}]{mailybaev2012}%
  \BibitemOpen
  \bibfield  {author} {\bibinfo {author} {\bibfnamefont {A.~A.}\ \bibnamefont
  {Mailybaev}},\ }\href@noop {} {\bibfield  {journal} {\bibinfo  {journal}
  {Phys. Rev. E}\ }\textbf {\bibinfo {volume} {85}},\ \bibinfo {pages} {066317}
  (\bibinfo {year} {2012}{\natexlab{a}})}\BibitemShut {NoStop}%
\bibitem [{\citenamefont {Gilson}\ and\ \citenamefont
  {Dombre}(1997)}]{gilson1997towards}%
  \BibitemOpen
  \bibfield  {author} {\bibinfo {author} {\bibfnamefont {J.~L.}\ \bibnamefont
  {Gilson}}\ and\ \bibinfo {author} {\bibfnamefont {T.}~\bibnamefont
  {Dombre}},\ }\href@noop {} {\bibfield  {journal} {\bibinfo  {journal} {Phys.
  Rev. Lett.}\ }\textbf {\bibinfo {volume} {79}},\ \bibinfo {pages} {5002}
  (\bibinfo {year} {1997})}\BibitemShut {NoStop}%
\bibitem [{\citenamefont {L'vov}\ \emph {et~al.}(2001)\citenamefont {L'vov},
  \citenamefont {Pomyalov},\ and\ \citenamefont {Procaccia}}]{l2001outliers}%
  \BibitemOpen
  \bibfield  {author} {\bibinfo {author} {\bibfnamefont {V.~S.}\ \bibnamefont
  {L'vov}}, \bibinfo {author} {\bibfnamefont {A.}~\bibnamefont {Pomyalov}}, \
  and\ \bibinfo {author} {\bibfnamefont {I.}~\bibnamefont {Procaccia}},\
  }\href@noop {} {\bibfield  {journal} {\bibinfo  {journal} {Phys. Rev. E}\
  }\textbf {\bibinfo {volume} {63}},\ \bibinfo {pages} {056118} (\bibinfo
  {year} {2001})}\BibitemShut {NoStop}%
\bibitem [{\citenamefont {Eggers}\ and\ \citenamefont
  {Grossmann}(1991)}]{eggers1991does}%
  \BibitemOpen
  \bibfield  {author} {\bibinfo {author} {\bibfnamefont {J.}~\bibnamefont
  {Eggers}}\ and\ \bibinfo {author} {\bibfnamefont {S.}~\bibnamefont
  {Grossmann}},\ }\href@noop {} {\bibfield  {journal} {\bibinfo  {journal}
  {Phys. Fluids A}\ }\textbf {\bibinfo {volume} {3}},\ \bibinfo {pages} {1958}
  (\bibinfo {year} {1991})}\BibitemShut {NoStop}%
\bibitem [{\citenamefont {Uhlig}\ and\ \citenamefont
  {Eggers}(1997{\natexlab{a}})}]{uhlig1997local}%
  \BibitemOpen
  \bibfield  {author} {\bibinfo {author} {\bibfnamefont {C.}~\bibnamefont
  {Uhlig}}\ and\ \bibinfo {author} {\bibfnamefont {J.}~\bibnamefont {Eggers}},\
  }\href@noop {} {\bibfield  {journal} {\bibinfo  {journal} {Z. Phys. B Con.
  Mat.}\ }\textbf {\bibinfo {volume} {102}},\ \bibinfo {pages} {513} (\bibinfo
  {year} {1997}{\natexlab{a}})}\BibitemShut {NoStop}%
\bibitem [{\citenamefont {Uhlig}\ and\ \citenamefont
  {Eggers}(1997{\natexlab{b}})}]{uhlig1997singularities}%
  \BibitemOpen
  \bibfield  {author} {\bibinfo {author} {\bibfnamefont {C.}~\bibnamefont
  {Uhlig}}\ and\ \bibinfo {author} {\bibfnamefont {J.}~\bibnamefont {Eggers}},\
  }\href@noop {} {\bibfield  {journal} {\bibinfo  {journal} {Z. Phys. B Con.
  Mat.}\ }\textbf {\bibinfo {volume} {103}},\ \bibinfo {pages} {69} (\bibinfo
  {year} {1997}{\natexlab{b}})}\BibitemShut {NoStop}%
\bibitem [{\citenamefont {Siggia}(1978)}]{siggia1978model}%
  \BibitemOpen
  \bibfield  {author} {\bibinfo {author} {\bibfnamefont {E.~D.}\ \bibnamefont
  {Siggia}},\ }\href@noop {} {\bibfield  {journal} {\bibinfo  {journal} {Phys.
  Rev. A}\ }\textbf {\bibinfo {volume} {17}},\ \bibinfo {pages} {1166}
  (\bibinfo {year} {1978})}\BibitemShut {NoStop}%
\bibitem [{\citenamefont {Nakano}(1988)}]{nakano1988}%
  \BibitemOpen
  \bibfield  {author} {\bibinfo {author} {\bibfnamefont {T.}~\bibnamefont
  {Nakano}},\ }\href@noop {} {\bibfield  {journal} {\bibinfo  {journal} {Prog.
  Theor. Phys.}\ }\textbf {\bibinfo {volume} {79}},\ \bibinfo {pages} {569}
  (\bibinfo {year} {1988})}\BibitemShut {NoStop}%
\bibitem [{\citenamefont {L'vov}(2002)}]{l2002quasisolitons}%
  \BibitemOpen
  \bibfield  {author} {\bibinfo {author} {\bibfnamefont {V.~S.}\ \bibnamefont
  {L'vov}},\ }\href@noop {} {\bibfield  {journal} {\bibinfo  {journal} {Phys.
  Rev. E}\ }\textbf {\bibinfo {volume} {65}},\ \bibinfo {pages} {026309}
  (\bibinfo {year} {2002})}\BibitemShut {NoStop}%
\bibitem [{\citenamefont
  {Mailybaev}(2012{\natexlab{b}})}]{mailybaev2012computation}%
  \BibitemOpen
  \bibfield  {author} {\bibinfo {author} {\bibfnamefont {A.~A.}\ \bibnamefont
  {Mailybaev}},\ }\href@noop {} {\bibfield  {journal} {\bibinfo  {journal}
  {Phys. Rev. E}\ }\textbf {\bibinfo {volume} {86}},\ \bibinfo {pages} {025301}
  (\bibinfo {year} {2012}{\natexlab{b}})}\BibitemShut {NoStop}%
\bibitem [{\citenamefont {Daumont}\ \emph {et~al.}(2000)\citenamefont
  {Daumont}, \citenamefont {Dombre},\ and\ \citenamefont
  {Gilson}}]{daumont2000instanton}%
  \BibitemOpen
  \bibfield  {author} {\bibinfo {author} {\bibfnamefont {I.}~\bibnamefont
  {Daumont}}, \bibinfo {author} {\bibfnamefont {T.}~\bibnamefont {Dombre}}, \
  and\ \bibinfo {author} {\bibfnamefont {J.~L.}\ \bibnamefont {Gilson}},\
  }\href@noop {} {\bibfield  {journal} {\bibinfo  {journal} {Phys. Rev. E}\
  }\textbf {\bibinfo {volume} {62}},\ \bibinfo {pages} {3592} (\bibinfo {year}
  {2000})}\BibitemShut {NoStop}%
\bibitem [{\citenamefont {G{\"a}rtner}(1977)}]{gartner1977large}%
  \BibitemOpen
  \bibfield  {author} {\bibinfo {author} {\bibfnamefont {J.}~\bibnamefont
  {G{\"a}rtner}},\ }\href@noop {} {\bibfield  {journal} {\bibinfo  {journal}
  {Theor. Probab. Appl.}\ }\textbf {\bibinfo {volume} {22}},\ \bibinfo {pages}
  {24} (\bibinfo {year} {1977})}\BibitemShut {NoStop}%
\bibitem [{\citenamefont {Ellis}(1984)}]{ellis1984large}%
  \BibitemOpen
  \bibfield  {author} {\bibinfo {author} {\bibfnamefont {R.}~\bibnamefont
  {Ellis}},\ }\href@noop {} {\bibfield  {journal} {\bibinfo  {journal} {Ann.
  Probab.}\ }\textbf {\bibinfo {volume} {12}},\ \bibinfo {pages} {1} (\bibinfo
  {year} {1984})}\BibitemShut {NoStop}%
\bibitem [{\citenamefont {Touchette}(2009)}]{touchette2009large}%
  \BibitemOpen
  \bibfield  {author} {\bibinfo {author} {\bibfnamefont {H.}~\bibnamefont
  {Touchette}},\ }\href@noop {} {\bibfield  {journal} {\bibinfo  {journal}
  {Phys. Rep.}\ }\textbf {\bibinfo {volume} {478}},\ \bibinfo {pages} {1}
  (\bibinfo {year} {2009})}\BibitemShut {NoStop}%
\bibitem [{\citenamefont {Richardson}(2007)}]{richardson2007weather}%
  \BibitemOpen
  \bibfield  {author} {\bibinfo {author} {\bibfnamefont {L.}~\bibnamefont
  {Richardson}},\ }\href@noop {} {\emph {\bibinfo {title} {{Weather Prediction
  by Numerical Process}}}}\ (\bibinfo  {publisher} {Cambridge University
  Press},\ \bibinfo {year} {2007})\BibitemShut {NoStop}%
\end{thebibliography}%

\end{document}